\documentclass[superscriptaddress, reprint, amsmath, amssymb, aps, prl]{revtex4-2}

\usepackage{graphicx}
\usepackage{dcolumn}
\usepackage{bm}
\usepackage{gensymb}
\usepackage[T1]{fontenc}
\usepackage{upgreek}
\usepackage{braket}
\usepackage{svg}
\usepackage{dsfont}
\usepackage[draft]{hyperref}

\renewcommand{\figurename}{Fig.}

\renewcommand{\Re}{\operatorname{Re}}
\renewcommand{\Im}{\operatorname{Im}}

\setcitestyle{super}

\begin{document}

\title{Interband Berry connection measurement in the optical honeycomb lattice}

\author{Shao-Wen Chang}
 \thanks{These authors contributed equally to this work.}
 \affiliation{Department of Physics, University of California, Berkeley, Berkeley, CA 94720, USA}
 \affiliation{Challenge Institute for Quantum Computation, University of California, Berkeley, Berkeley, CA 94720, USA}

\author{Malte N. Schwarz}%
 \thanks{These authors contributed equally to this work.}
 \affiliation{Department of Physics, University of California, Berkeley, Berkeley, CA 94720, USA}
 \affiliation{Challenge Institute for Quantum Computation, University of California, Berkeley, Berkeley, CA 94720, USA}
 
\author{Erin G. Moloney}%
 \affiliation{Department of Physics, University of California, Berkeley, Berkeley, CA 94720, USA}
 \affiliation{Challenge Institute for Quantum Computation, University of California, Berkeley, Berkeley, CA 94720, USA}
 
\author{Ke Lin}%
 \affiliation{Department of Physics, University of California, Berkeley, Berkeley, CA 94720, USA}
 \affiliation{Challenge Institute for Quantum Computation, University of California, Berkeley, Berkeley, CA 94720, USA}
 
\author{Dan M. Stamper-Kurn}%
 \email{dmsk@berkeley.edu}
 \affiliation{Department of Physics, University of California, Berkeley, Berkeley, CA 94720, USA}
 \affiliation{Challenge Institute for Quantum Computation, University of California, Berkeley, Berkeley, CA 94720, USA}
 \affiliation{Materials Sciences Division, Lawrence Berkeley National Laboratory, Berkeley, CA 94720, USA}

\date{\today}
\begin{abstract}
    The geometry of Bloch bands affects many physical properties of crystalline solids and other spatially periodic systems.  Direct experimental determination of such geometry is an active area of research.  In this work, we focus on the fundamental connection between optical excitations and the relative geometry of pairs of Bloch bands, as characterized by the interband Berry connection.  We simulate the response of electrons in solids to optical excitation by the response of ultracold fermionic atoms in optical lattices to periodic modulation of the lattice position.  The strength of resonant excitation between bands, measured at each quasimomentum and for various lattice-shaking polarizations, directly maps out the interband Berry connection.  We apply this method to the optical honeycomb lattice, driving excitations between the ground $n=1$ band and the excited $n'=\{2,3,4\}$ bands.  We observe transparency lines of quasimomenta at which the response to excitation of specific polarization is zero.  Further, the interband Berry connection between bands 1 and 3 shows irreducible Dirac strings connecting the $K$ and $K'$ points in the Brillouin zone, lines along which the interband Berry connection abruptly changes orientation.   Our work establishes optical response as a powerful tool for characterizing geometrical and topological properties of band structure.
\end{abstract}

\maketitle

The energy eigenstates of particles moving within spatially periodic structures make up a series of Bloch bands, each a continuous manifold of states parametrized by the quasimomentum $\vec{q}$.  Connections have been established between geometric properties of these Bloch-band manifolds and  physical properties of materials.  Examples include the relations between single-band geometry and the anomalous Hall effect in itinerant ferromagnets \cite{jung02anomalous}, the motion of wavepackets \cite{chan96berry,jotz14haldane}, and electric polarization \cite{rest94rmp}. Summing up the geometry of bands over the  Brillouin zone gives topological invariants that are also found to relate to physical properties of materials, such as the edge modes of Chern insulators \cite{qi_topological_2011}.

One measure of the geometry of Bloch bands is the Berry connection $\vec{A}_{n n'} (\vec{q}\,) = i \braket{u_{n \vec{q}}\,|\nabla_{\vec{q}}|u_{n' \vec{q}}}$, where $\ket{n \vec{q}} =  e^{i \vec{q}\cdot\hat{\vec{r}}} \ket{u_{n \vec{q}}}$ are the Bloch states and $\ket{u_{n \vec{q}}}$ are the periodic unit-cell wavefunctions.  The diagonal Berry connection, with band indices $n = n'$, characterizes the geometry of a single Bloch band.  Its closed-loop line integrals give the Abelian Berry phase acquired through parallel transport within the band \cite{berr84phase, xiao10rmp}.  The off-diagonal Berry connection, with $n \neq n'$, characterizes the relative geometry of two Bloch-band manifolds \cite{ma_abelian_2010}. Line integrals of the interband Berry connection yield the non-Abelian Wilson loop operators \cite{wilc84nonabelian}.

Here, we focus on the essential relation between band geometry  and the optical response of solids.  The strength of an allowed transition from band $n$ to $n'$ is determined by the off-diagonal matrix elements of the position operator $\hat{\vec{r}}$, given as \cite{blou62,rest94rmp}
\begin{equation}
    \bra{n \vec{q}\,} \hat{\vec{r}} \ket{n' \vec{q}\,'}  = i \delta(\vec{q}-\vec{q}\,') \vec{A}_{n n'}(\vec{q}\,)
\end{equation}
Optical transitions in solids drive electrons between bands with nearly no change in quasimomentum, i.e., $\vec{q} = \vec{q}\,'$.  For a weak pulsed optical drive, the probability of resonant electronic excitation between bands at quasimomentum $\vec{q}$ then becomes
\begin{equation}
    P_{n n'}(\vec{q}) = \frac{|F|^2}{\hbar^2} |\vec{\epsilon} \cdot \vec{A}_{n n'}(\vec{q})|^2 t^2
    \label{eq:rate}
\end{equation}
where $\vec{\epsilon}$ is the optical polarization vector, the force magnitude $F$ is given by the product of the electron charge and the optical electric field magnitude, and $t$ is the pulse duration. Thus, the optical response strength stems directly from the interband Berry connection, i.e.\ from the relative geometry of two bands connected by an optical transition.  This geometric interpretation is relevant to photoconductivity \cite{jankowski_optical_2025, onishi_fundamental_2024},
nonlinear optical response \cite{hughes_calculation_1996, morimoto_topological_2016}, the photovoltaic effect \cite{dai_recent_2023, de_juan_quantized_2017}, and optical selection rules for excitons \cite{cao_unifying_2018}. 

In this work, we simulate the response of electrons in solids to optical excitation by the response of ultracold fermionic atoms in optical lattices to periodic modulation of the lattice position. Whereas the optical response in solids is typically observed through bulk measures, such as photoconductivity or optical absorption, that sum over the response of electrons at all quasimomenta, in our ultracold-atom quantum simulator, we resolve the resonant response of atoms separately at each quasimomentum.  Thus, we are able to evaluate the excitation probability for various lattice-shaking polarizations $\vec{\epsilon}$, utilize Eq.~\ref{eq:rate}, and map out the interband Berry connection $\vec{A}_{n n'}(\vec{q}\,)$ across the Brillouin zone. We apply this method to observe directly the interband Berry connections between the ground $n=1$ and the excited $n' = \{2,3,4\}$ bands of the two-dimensional honeycomb lattice. Our measurements reveal detailed features of these vector fields, including ``transparency lines'' of quasimomenta where the response to excitation of specific polarizations is zero, and irreducible Dirac strings (in $\vec{A}_{13}$) connecting the $K$ and $K'$ points in the Brillouin zone, lines along which the interband Berry connection abruptly changes orientation.

We follow several preceding works on characterizing band geometry  from experimental data. The Hall response of ultracold atoms was used to characterize single-band geometry and topology in static and Floquet-driven optical lattices \cite{jotz14haldane, aide15chern,wint20floquet}.  Parallel transport and interferometry of ultracold atoms were used to characterize the Abelian and non-Abelian Berry connections near band-touching points \cite{duca15ab,brow22singularity}. State-tomography methods were used to reconstruct the unit-cell wavefunctions across the Brillouin zone in solids \cite{kim25qmt}, exciton-polaritons \cite{gianfrate_measurement_2020}, optical lattices \cite{flas16Berry,li16wilson,yi23qgt}, plasmonic lattices \cite{cuerda_observation_2024} and photonic crystals \cite{guil25quantumgeometry}.  These state reconstructions can be differentiated numerically to determine the Berry connection and the complete quantum geometric tensor. In comparison, the present work highlights and exploits the direct physical connection between the response of materials to optical excitation (or, rather, the optical-lattice equivalent of lattice shaking) and the interband Berry connection. Closely related to our work, the response of an atomic gas to circular shaking, integrated over the entire Brillouin zone, was used to extract the Chern number of one band of a Floquet-driven optical lattice \cite{tran_probing_2017, asteria_measuring_2019}.

\begin{figure}
    \centering
    \includegraphics[width=90mm]{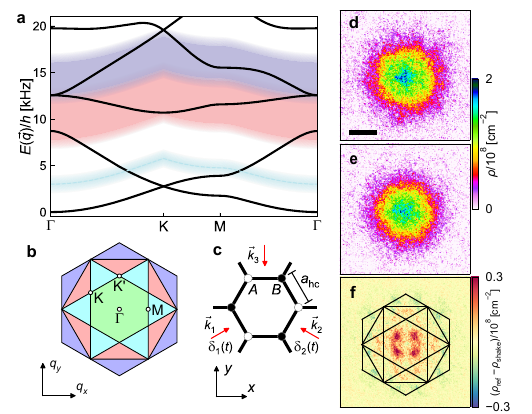}
    \caption{\textbf{Experiment setup.}
        \textbf{a}, Band structure of the honeycomb lattice at $V_0 = 8.95\, E_R$. For excitation from the ground band, the blue dashed line corresponds to the $3\,$kHz shaking frequency used for Fig.~\ref{fig:2:angle_scan}. Red and purple shaded regions correspond to  frequency ranges to measure $\vec{A}_{13}$ and $\vec{A}_{14}$, respectively.  The Fourier width of the finite shaking time is represented by a $1\,$kHz wide gradient. 
        \textbf{b}, The first four Brillouin zones of the honeycomb lattice with high symmetry points labeled.
        \textbf{c}, Illustration of the honeycomb lattice formed by three $1064\,$nm wavelength beams (red arrows).
        \textbf{d}-\textbf{f}, Quasimomentum-resolved excitation for linear shaking along $x$ at $10.86\,$kHz.
     Single-shot images show atoms after (\textbf{d}, $\rho_\mathrm{shake}$) and without (\textbf{e}, $\rho_\mathrm{ref}$) shaking. \textbf{f}, Difference image $\rho_\mathrm{ref}-\rho_\mathrm{shake}$ averaged over 14 repetitions.   The first four Brillouin zones are drawn in black lines.  Scale bar for \textbf{d}-\textbf{f} is 0.1 mm.}
    \label{fig:1:experiment_scheme}
\end{figure}
We begin our experiment with a single-component degenerate Fermi gas of around $8 \times 10^4$ $^{40} \mathrm{K}$ atoms trapped optically (see \hyperref[sec:methods]{Methods}). A static honeycomb optical lattice with lattice spacing $a_{\mathrm{hc}} = 410\,$nm is formed using three light beams (wavelength $\lambda = 1064$ nm) that propagate in the horizontal $x$-$y$ plane and intersect at equal angles. We adiabatically ramp the lattice on to a final depth $V_0 = 8.95\,E_\mathrm{R}$ in $150\,$ms, and keep it at $V_0$ for an additional $100\,$ms. Here, $E_\mathrm{R} = h^2 / 2 m \lambda^2 = h \times 4.41\,$kHz is the recoil energy, with $m$ being the atomic mass.  The chemical potential is such that the $n=1$ band is well filled, the $n=2$ band partly filled, and higher bands empty.

To probe the system, we periodically modulate the relative phases of lattice beam pairs by controlling the frequency detunings of two of the lattice beams, $\delta_1(t)$ and $\delta_2(t)$ (Fig.~\ref{fig:1:experiment_scheme}c). In the lab frame, the velocity of the lattice is then given by 
$\frac{2}{3} \lambda (\delta_1(t) \vec{k}_1 + \delta_2(t) \vec{k}_2) / k_\mathrm{lat}$, where the lattice beam wave vectors $\vec{k}_i$, each with magnitude $k_\mathrm{lat} = 2 \pi/\lambda$, are oriented as shown in Fig.\ \ref{fig:1:experiment_scheme}c.
In a frame co-moving with the optical lattice, lattice shaking adds a perturbation of the form
\begin{align}
    \hat{H'} = 2 F\, \Re[e^{-i \omega_{\mathrm{PM} } t} \vec{\epsilon}\,] \cdot \hat{\vec{r}},
\end{align}
where $F$ is the shaking force amplitude, $\vec{\epsilon}$ is the normalized shaking polarization vector, lying within the $x$-$y$ plane, and $\omega_\mathrm{PM} = 2\pi f_{\textrm{PM}}$ is the angular frequency of the drive. We then obtain Eq.~\eqref{eq:rate} within first-order perturbation theory. For all the results presented in the main text, we chose $F / (E_\mathrm{R} / a_{\mathrm{hc}}) = 0.38$.
The shaking is applied for approximately $1\,$ms (see \hyperref[sec:methods]{Methods} for details on pulse shaping).

After shaking, the lattice is turned off abruptly and the atoms propagate freely for $14\,$ms before being imaged. We quantify the excitation probability from band $n=1$ to higher bands $n'$ through the fractional depletion of atomic population, $D_{n n'}(\vec{q}) = (\rho_\mathrm{ref}-\rho_\mathrm{shake})/\rho_\mathrm{ref}$, where the column densities with ($\rho_\mathrm{shake}$) and without shaking ($\rho_\mathrm{ref}$) are evaluated only within the first Brillouin zone. We infer $n'$ from the shaking frequency. This depletion $D_{n n'}(\vec{q})$ is systematically smaller than the population excitation probability $P_{n n'}(\vec{q})$ because of the distribution of Bloch-state populations across several Brillouin zones and initial population in excited bands (see \hyperref[sec:diff_and_BM]{Methods} for further discussion).

\begin{figure}
    \centering
    \includegraphics[width=90mm]{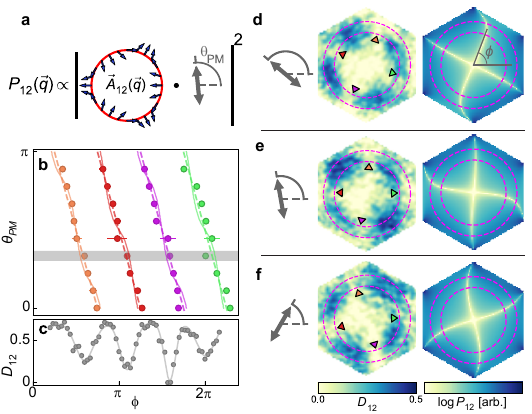}
    \caption{\textbf{Polarization-dependent optical excitation between the lowest two bands.}
    \textbf{a}, Graphical representation of the matrix element. The direction of $\vec{A}_{21}$ is plotted along a circle around the $\mathrm{\Gamma}$ point. The direction of linear shaking along $\theta_{\mathrm{PM}}$ is indicated by the gray arrow. Transparency at $\vec{q}$ occurs when $\vec{A}_{12} \cdot \vec{\epsilon}(\theta_{\mathrm{PM}}) = 0$.
    \textbf{b}, Angular positions $\phi$ of the transparency lines as a function of $\theta_{\mathrm{PM}}$. The positions are obtained by fitting a sum of four Gaussian functions to the angular integral of the excitation ratio within the annular region enclosed in the purple dashed lines in panels \textbf{d}~-~\textbf{f}.  An example of such a fit for one $\theta_\mathrm{PM}$ (gray shade in \textbf{b}) is shown in \textbf{c}.  Error bars for $\phi$ are two standard deviations of the fitted positions. Dashed lines are predictions from a two-band tight-binding model (see \hyperref[sec:tight_binding]{Methods}). Solid lines are from numerical calculations for the experimental lattice depth, evaluated at $|\vec{q}\,| = 0.66\, |\vec{k}|$.
    \textbf{d}~-~\textbf{f}, Measured $D_{12}$  for $\theta_{\mathrm{PM}} = (7\pi/9,\,5\pi/9,\,3\pi/9)$ in (\textbf{d}, \textbf{e},~\textbf{f}), respectively. Predicted values of $\log P_{12}$, for the same shaking excitation, are shown next to the experiment data for comparison. Logarithm is taken to exaggerate the transparency lines.
    }
    \label{fig:2:angle_scan}
\end{figure}

As an example of how the interband Berry connection affects physical observables, we first focus on the lowest two bands of the honeycomb lattice. From Eq.~\eqref{eq:rate}, one sees that the excitation rate vanishes when the matrix element of the perturbation is zero, which happens when $\vec{\epsilon} \perp \vec{A}_{nn'}(\vec{q}\,)$.
In this sense, one can say that the direction of the vector field $\vec{A}_{nn'}(\vec{q})$ imposes selection rules for optical excitations at quasimomentum $\vec{q}$. We study this effect by shaking the lattice linearly at a fixed shaking frequency $\omega_{\mathrm{PM}} = 2 \pi \times 3\,$kHz and with the shaking direction rotated by an angle $\theta_\mathrm{PM}$ with respect to the $x$-axis.

The measured excitation ratio is plotted in Fig.~\ref{fig:2:angle_scan}, which shows clear signs of the transparency lines with $\theta_{\mathrm{PM}}$-dependent positions. This behavior can be explained within a two-band tight-binding model. An expansion around $\Gamma$ predicts that there are four values of $\phi$ with vanishing excitation, with their positions changing by $- \theta_{\mathrm{PM}} / 2$ as the shaking direction varies. This behavior is captured by the experiment, with the angular positions of the minima $\phi$ agreeing to within 0.25 radians. This measurement reveals how the double winding of the field $\vec{A}_{12}(\vec{q})$ about the $\Gamma$ point affects optical excitation properties, even though the bands involved in the transition show no particular features such as band singularity around $\Gamma$.

\begin{figure}
    \centering
    \includegraphics[width=90mm]{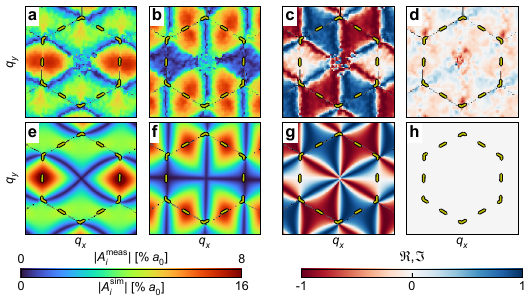}
    \caption{\textbf{Interband Berry connection $\vec{A}_{14}$.}
    \textbf{a},~\textbf{b}, Measured vector field projection magnitudes $|A_x|$(\textbf{a}) and $|A_y|$(\textbf{b}) extracted from $D_{14}$. Yellow dashed lines label the first Brillouin zone.  The data are repeated in an extended zone scheme. Transparency lines in \textbf{a} (\textbf{b}) correspond to quasimomenta where the projection of $\vec{A}$ along $x$ ($y$) is zero.
    \textbf{c},~\textbf{d}, The normalized real part $\mathfrak{R}$ (\textbf{c}) and imaginary part $\mathfrak{I}$ (\textbf{d}) of the cross term $(2 A_x A_y^*) / (|A_x|^2 + |A_y|^2)$.
    \textbf{e}~-~\textbf{h}, Simulations of $|A_x|$  (\textbf{e}), $|A_y|$  (\textbf{f}) and the normalized real (\textbf{g}) and imaginary (\textbf{h}) part of the normalized cross term in a $8.95 E_R$ deep lattice calculated using a plane wave expansion. The simulation is based on $P_{14}$ and the quantitative difference to the measured data is described in the main text.}
    \label{fig:3:Full_14_interband_Berry_connection}
\end{figure}

The region where we can measure the excitation between the lowest two bands reliably is limited by the fact that the band gaps near $K$ and $K'$ approach zero. To demonstrate our capability to measure the structure of $\vec{A}_{nn'}$ across the entire Brillouin zone, we therefore turn to transitions from the ground band to higher energy bands. We shake the lattice linearly at $\theta_{\text{PM}} = 0\degree, 45\degree, 90\degree$ and $135\degree$, and also circularly with both handednesses.   We determine the interband resonance shaking frequency separately at each quasimomentum, and use the measured depletion fractions at that frequency, at all applied polarizations, to determine the interband Berry connection, as discussed below  (see \hyperref[sec:methods]{Methods} for more details).

\begin{figure*}[t]
    \begin{minipage}{0.999\textwidth}
        \centering
        \includegraphics[width=1\linewidth]{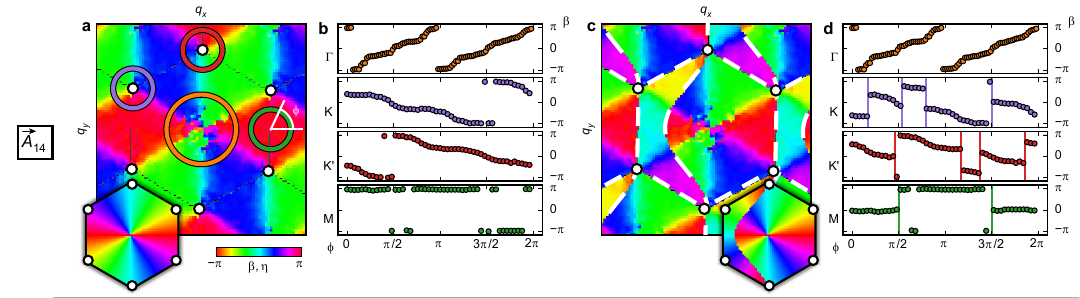}
    \end{minipage}

    \begin{minipage}{0.999\textwidth}
        \centering
        \includegraphics[width=1\linewidth]{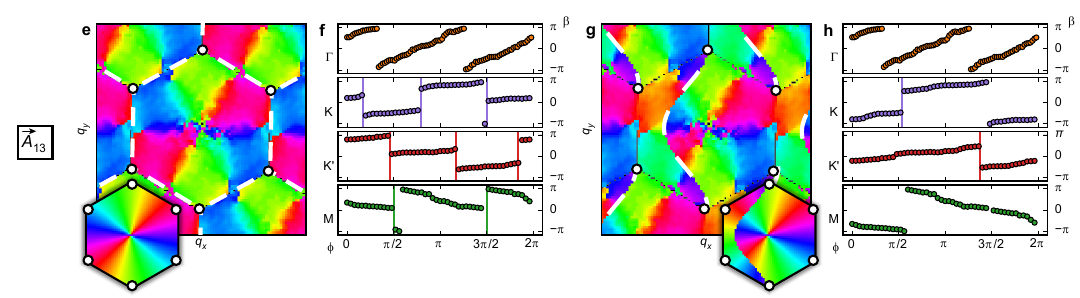}
    \end{minipage}
 
    \caption{\textbf{The orientation $\beta$ of $\vec{A}_{14}$ and $\vec{A}_{13}$ in different gauges.}
    \textbf{a},~\textbf{b}, $\beta$ for $\vec{A}_{14}$ in a continuous gauge choice exhibits no phase jumps away from high-symmetry points. Unfilled circles in \textbf{a} indicate paths taken in \textbf{b}. The hexagonal inset shows the gauge map $\eta(\vec{q})$ used to evaluate $\beta = \arctan( \textrm{sgn}(\mathfrak{R}) |A_y| / |A_x| )$.  White-filled circles in both figure and inset indicate corners of the Brillouin zone. 
    \textbf{c},~\textbf{d}, By changing the gauge choice  (see inset), lines of discontinuity in $\beta$ are introduced, marked as white dashed lines; these are Dirac strings.  On closed paths circumscribing high-symmetry points (\textbf{d}) $\pi$-valued phase jumps in $\beta$ occur upon crossing Dirac strings.  Here, these crossings occur in pairs. The parity indicates that these Dirac strings can be gauged away (as they are in \textbf{a}).
    \textbf{e},~\textbf{f}, $\beta$ for $\vec{A}_{13}$ under a gauge map $\eta(\vec{q})$ that is continuous inside the Brillouin zone. Dirac strings connecting $K$ and $K'$ points occur at the Brillouin zone edges. Where the orientation $\beta$ on closed paths (\textbf{f}) reveals an odd number of $\pi$-valued phase jumps, a gauge choice cannot remove all Dirac strings. \textbf{g},~\textbf{h}, A different gauge (see inset) moves the Dirac strings connecting the $K$ and $K'$ points to lie within the Brillouin zone.}
    \label{fig:4:Dirac_string}
\end{figure*}
Measurements of the interband Berry connection $\vec{A}_{14}$ determined by this method are shown in Fig.~\ref{fig:3:Full_14_interband_Berry_connection} (similar data for $\vec{A}_\mathrm{13}$ are shown in Extended Data Fig.~\ref{fig:FigA_A13}).  We observe distinct transparency lines in both $|A_x|$ and $|A_y|$ traversing the entire Brillouin zone.
The transparency lines for both polarizations are anchored to $\Gamma$ and $K$, indicating that the interband Berry connection vanishes there.

Our measurements reveal not only the magnitudes $|A_x|$ and $|A_y|$ of projections of the interband Berry connection, but also the gauge-invariant relative phase of these projections. To characterize this relative phase, we fit our excitation data to determine the normalized cross term $(2 A_x A_y^*) / (|A_x|^2 + |A_y|^2)$. The real part of the normalized cross term, $\mathfrak{R}$, which depends on the choice of linear basis, shows the intricate structure of the relative sign between the two projections $A_x$ and $A_y$. The imaginary part of the normalized cross term, $\mathfrak{I}$, invariant under basis rotations,
specifies the quasimomentum-dependent circular dichroism of the system. The honeycomb lattice exhibits time-reversal symmetry, implying that no circular dichroism should appear, and, indeed, we measure a vanishing $\mathfrak{I}$, up to measurement imperfections.

Comparing our measurements to numerical simulations based on non-interacting particle dynamics in our honeycomb optical lattice (Fig.\ \ref{fig:3:Full_14_interband_Berry_connection}), we see that the experiment faithfully reveals many fine features of the interband Berry connection across the Brillouin zone. However, we find that the measured amplitude of $\vec{A}$ is systematically smaller, by about a factor of two, from the theoretical results. We attribute this underestimation primarily to two effects. First, holes created by excitations evolve dynamically in quasimomentum during the modulation, ``smearing out'' the response over the Brillouin zone \cite{heinze_intrinsic_2013}. Second, at certain quasimomenta, primarily close to zone edges, our imaging procedure does not fully differentiate atomic population in neighboring Brillouin zones (see \hyperref[sec:methods]{Methods}). 
Still, the ability to detect the position of transparency lines highlights the ability of our measurement technique to  extract precisely the direction, if not the magnitude, of the vector field. 

We have focused thus far on extracting gauge-invariant quantities from measurements of the lattice-shaking response of a quantum gas. Converting these measurements to a full map of the interband Berry connection across the Brillouin zone requires that we fix a gauge.  Having determined experimentally that the ratio $A_y/A_x$ is real, we may express the Berry connection as 
$\vec{A} = A_r (\cos(\beta) \vec{e}_x + \sin(\beta) \vec{e}_y)$, where $A_r = \sqrt{|A_x|^2 + |A_y|^2}$, $\vec{e}_{x,y}$ are unit vectors along the $x$ and $y$ axis, respectively, and $\beta = \arctan( \textrm{sgn}(\mathfrak{R}) |A_y| / |A_x| )$ is the vector orientation in the $x$-$y$ plane.
Gauge fixing is now equivalent to choosing a $\pi$ interval range for the arctan function at each quasimomentum.  We specify this range as the interval $\beta \in [ \eta(\vec{q}) - \pi/2, \eta(\vec{q}) + \pi/2]$.

The orientations $\beta(\vec{q})$ of the vector fields $\vec{A}_{14}(\vec{q})$ and $\vec{A}_{13}(\vec{q})$, determined from experimental data, are shown in Fig.\ \ref{fig:4:Dirac_string}, each for two different choices of gauge $\eta(\vec{q})$.  We find that one can choose a gauge in which the orientation of the interband Berry connection $\vec{A}_{14}(\vec{q})$ varies smoothly at all quasimomenta (Fig.~\ref{fig:4:Dirac_string}a,b).  We highlight this smooth variation by tracing the value of $\beta$ along circular paths in quasimomentum that circumscribe high-symmetry points in the Brillouin zone.

In contrast, we find that, regardless of the choice of gauge, the interband Berry connection $\vec{A}_{13}(\vec{q})$ always contains loci along which the orientation of $\vec{A}$ abruptly changes sign.  We identify these loci as Dirac strings that connect $K$ and $K'$ points in the Brillouin zone. For example, in Fig.\ \ref{fig:4:Dirac_string}g and h, with a particular choice of gauge, we find that the orientation $\beta$ shows one $\pi$-phase jump along circular paths that encircle either the $K$ or $K'$ point, as that path encounters a single Dirac string.
The positions and number of Dirac strings vary with different gauge choices. For example, the discontinuous gauge choice represented in Fig.~\ref{fig:4:Dirac_string}c introduces Dirac strings to $\vec{A}_{14}$. Paths around $K$ and $K'$ now cross four Dirac strings each. Similarly, the gauge choice shown in Fig.~\ref{fig:4:Dirac_string}e places Dirac strings in $\vec{A}_{13}$ along the Brillouin zone boundaries, so that paths around $K$ and $K'$ now cross three Dirac strings each. In both cases, the parity of the number of Dirac-string crossings along closed paths in quasimomentum is conserved.

In summary, we show that optical responses are powerful probes for the quantum geometry of band structures through their relation to the interband Berry connection. Our realization of such measurements in a cold atom system reveals quasimomentum-dependent optical selection rules in honeycomb lattices. The vector interband Berry connection field over the entire Brillouin zone is measured and agrees well with theory, up to a quasimomentum-dependent scaling factor. We observe an irreducible Dirac string in the interband Berry connection between the ground band and second excited band. As an extension of the method pursued in this work, by integrating the optical response over all interband transitions one can experimentally determine the full quantum metric tensor \cite{ozawa_extracting_2018}.

Our work paves the way towards studying non-trivial multi-band topology in geometrically frustrated flat band systems, such as the kagome lattice \cite{jo12kag}. The real interband Berry connection of neighboring bands, known as the Euler connection, carries information about topological frame-rotation charges in multi-band system. This quantity has been measured in ion trap systems \cite{zhao_quantum_2022}, but novel non-trivial fragile topological phases that are predicted to exist \cite{slager_non-abelian_2024, po_fragile_2018, unal_topological_2020} have not been observed. In the kagome lattice, the ground band manifold is predicted to have a non-zero Euler class around the QBTP of the flat band, ideal to test multi-band topological braiding schemes.

\section*{Acknowledgements}
We thank Johannes Mitscherling for discussions.  We acknowledge support from the NSF (grant number PHY-2309300 and the QLCI program through grant number OMA-2016245).  This work was performed in part at the Aspen Center for Physics, which is supported by National Science Foundation grant PHY-2210452, and we thank Daniel Arovas, Kenneth Burch, and Gil Raphael for discussions that took place there. S.-W.C.\ and M.N.S.\ are supported by Hearts to Humanity Eternal Association (H2H8). S.-W.C.\ is supported by the Ministry of Education, Republic of China (Taiwan). E.G.M.\ is supported by the Natural Sciences and Engineering Research Council of Canada (NSERC).

\section{Methods}\label{sec:methods}
\setcounter{figure}{0}
\renewcommand{\figurename}{Extended Data Fig.}
\subsection{Experimental setup}
In our setup, ultracold gas mixtures of $^{40}$K and $^{87}$Rb are produced by laser cooling, co-trapped in a plugged spherical-quadrupole trap and then an optical trap.   Forced evaporation of the $^{87}$Rb atoms produces quantum degenerate gases of either $^{40}$K, $^{87}$Rb, or both elements simultaneously.  In this paper, we produce a degenerate Fermi gas of around $8 \times 10^4$ $^{40}\mathrm{K}$ atoms at a temperature of $\sim0.2 \, T / T_\mathrm{F}$ in a crossed optical dipole trap by evaporating away all $^{87}$Rb atoms.
$T_F$ is the Fermi energy, which is around $h \times 10\;$kHz. We note that the average of the local Fermi energy is around $h \times 6\;$kHz, consistent with our observation that the third and higher bands are completely empty, and that the ground band is significantly more populated than the second band. The optical honeycomb lattice is formed by three phase-coherent light beams at $\lambda = 1064\,$nm, propagating at $120 \degree$ with respect to each other in the $x$-$y$-plane perpendicular to gravity. The resulting potential is given by
\begin{equation*}
    V(\vec{r}) = -\frac{2}{9} V_0 \left(3
- \cos(\vec{G}_1 \cdot \vec{r}) 
- \cos(\vec{G}_2 \cdot \vec{r})
- \cos(\vec{G}_3 \cdot \vec{r}) \right)
\end{equation*}
where $\vec{G}_1 = \vec{k}_2 - \vec{k}_3$ and $\vec{G}_2 = \vec{k}_3 - \vec{k}_1$ are primitive reciprocal lattice vectors, and $\vec{G}_3 = \vec{k}_1 - \vec{k}_2$. At the full lattice depth $V_0 = 8.95 \, E_\mathrm{R}$ used in this work, the lattice beams and trapping beams creates an approximately harmonic potential with trap frequencies $\omega_{x, y, z} = 2\pi \times (69,\,97,\,353)\,$Hz.

The data shown in the main text are taken with a total modulation time between $1.00$ and $1.06\,$ms. The modulation amplitude of lattice shaking is ramped up in $150\,\upmu$s, and then left at constant modulation amplitude. We enforce that the shaking ends with a non-moving lattice to avoid unwanted displacement in the final image. For linear shaking, this is done by choosing the initial phase of the modulation such that the lattice is stationary at the end. For circular shaking, since the speed of the lattice is constant at a given amplitude, we ramp down the modulation amplitude to zero in the last $150\,\upmu$s instead. When using Eq.~\eqref{eq:rate} to extract the value of interband Berry connection, we account for the ramp and the small steps in modulation time by using the pulse area for $t_{\mathrm{PM}}$.

The position of the lattice is feedback stabilized by a phase lock setup \cite{brow22singularity} whenever the lattice depth is larger than $0.05 E_R$. We confirm that the deviation from the programmed displacement is less than $a_{\text{hc}} / 10$ at any time during lattice shaking.

The quasimomentum resolution of our experiment is limited by the random walk of atom position as the atom scatters light from the imaging pulse during our absorption imaging sequence. The standard deviation of the distribution in final atom position after an $80\, \upmu$s imaging pulse with intensity $0.15\, I_\mathrm{sat}$ is $3.38\, \upmu$m, where $I_\mathrm{sat}$ is the saturation intensity. This deviation is comparable to the resolution of our imaging system, which is limited by the pixel size of the camera to be $3.7\, \upmu$m. We convolve each averaged image with a one-pixel wide normalized Gaussian mask before further analysis.

\begin{figure}
    \centering
    \includegraphics[width=90mm]{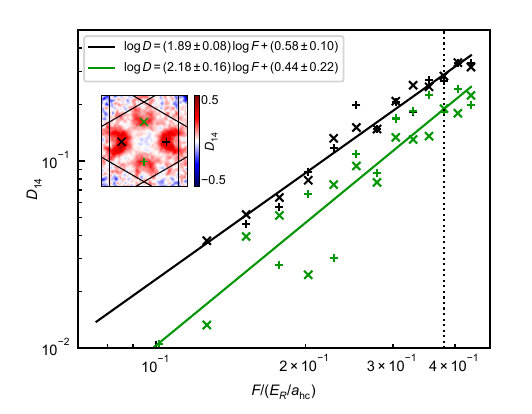}
    \caption{\textbf{Horizontal shaking at $14.35\,$kHz at different modulation strengths.}
    Each set of data is taken from the two parity-symmetric positions marked in the inset. Solid lines are fits to linear functions. The black dashed line labels $F/(E_R / a_\mathrm{hc}) = 0.38$ used in the main text. Inset: an example run at $F/(E_R / a_\mathrm{hc}) = 0.38$ with quasimomenta markers.}
    \label{fig:FigA_TDPT_Mscan}
\end{figure}
We restrict our analysis of the dynamics of the system using Eq.~\eqref{eq:rate}, which holds when the perturbation is close to the resonance. To confirm that a first-order perturbation theory treatment appropriately describes our measurement results, we perform two experiments at a representative modulation frequency $\omega_\mathrm{PM} = 2 \pi \times 14.35\,$kHz. We first fix $t_\mathrm{PM} = 1\,$ms and vary $0.08 \leq F / (E_R / a_\mathrm{hc}) \leq 0.43$. A log-log fit to $D_{14}$ as a function of $F$ gives scaling factors reasonably close to the expected quadratic scaling, see Extended Data Fig.~\ref{fig:FigA_TDPT_Mscan}.

\begin{figure}
    \centering
    \includegraphics[width=90mm]{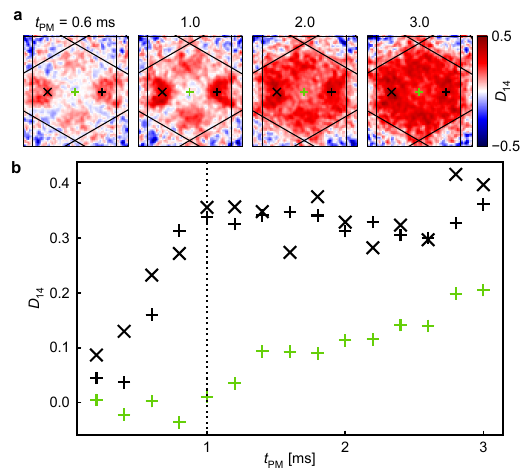}
    \caption{\textbf{Horizontal shaking at $14.35\,$kHz for varying durations.}
    \textbf{a} $D_{14}$ measurements for selected $t_\mathrm{PM}$ with quasimomenta markers for \textbf{b}.
    \textbf{b} $D_{14}$ as a function of $t_\mathrm{PM}$. Black dashed line labels $t_\mathrm{PM} = 1\,$ms used in the main text.}
    \label{fig:FigA_TDPT_tscan}
\end{figure}
In the second experiment, we shake at $F / (E_R / a_\mathrm{hc}) = 0.38$, while varying $t_\mathrm{PM}$ up to $3\,$ms. The measured excitation ratio deviates strongly from the expected quadratic scaling for both shaking strengths, especially at longer times, see Extended Data Fig.~\ref{fig:FigA_TDPT_tscan}. We emphasize that this deviation is not explained by two-level system dynamics, from which one expects Rabi-like oscillatory behavior. Instead, we attribute this to the dynamics of holes in the ground band \cite{heinze_intrinsic_2013}, which is supported by the observation that at longer times the measured excitation fraction becomes non-zero even at off-resonant $\vec{q}$.
\begin{figure*}
    \centering
    \includegraphics[width=1\linewidth]{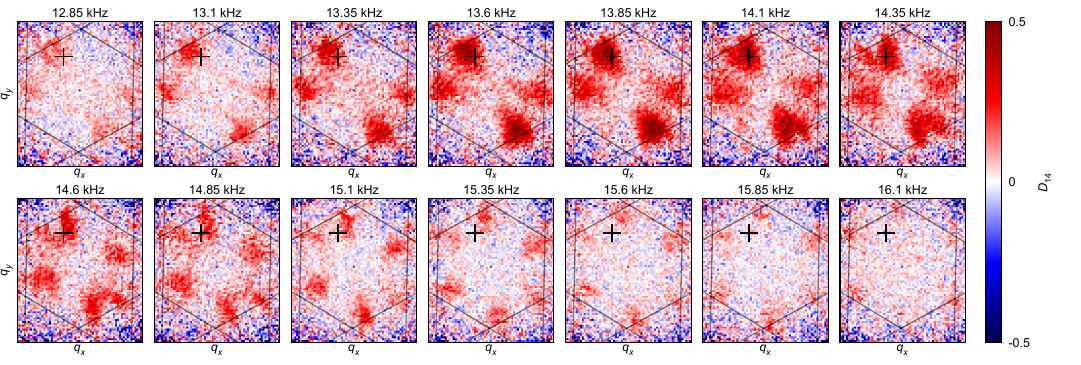}
    \caption{\textbf{The fractional depletion of atomic population, $D_{14}$, from diagonal ($\theta_\mathrm{PM}=45\degree, 225\degree$) linear shaking.}
    In the first Brillouin zone (hexagonal black outline), the atomic population depletes at different pixels, corresponding to different quasimomenta, when the shaking frequency (sub-figure labels) is resonant. The black cross marks the pixel analyzed in Extended Data Fig.~\ref{fig:FgA_maps_plus_spectrum}}.
    \label{fig:FigA_diffraction_panels}
\end{figure*}

\subsection{Tight binding model for the lowest two bands}\label{sec:tight_binding}
In tight-binding model, the momentum space Hamiltonian for the honeycomb lattice is given by
\begin{equation}
    H(\vec{q}) = -J
    \begin{pmatrix}
        0 & g(\vec{q})\\
        g^*(\vec{q}) & 0
    \end{pmatrix},
\end{equation}
where $g(\vec{q}) = \sum_i{e^{-i \vec{q} \cdot \vec{a}_i}}$, $J$ is the hopping strength, and $\vec{a}_i$'s are the vectors that connect one lattice $A$ site to the nearest-neighbor $B$ sites (see Fig.~\ref{fig:1:experiment_scheme}c). For the honeycomb lattice studied in this paper, we have
\begin{equation*}
    \vec{a}_1 = a_{\mathrm{hc}} \begin{pmatrix} 1 \\ 0 \end{pmatrix},\ 
    \vec{a}_2 = a_{\mathrm{hc}} \begin{pmatrix} - \frac{1}{2} \\ \frac{\sqrt{3}}{2} \end{pmatrix},\ 
    \vec{a}_3 = a_{\mathrm{hc}} \begin{pmatrix} - \frac{1}{2} \\ - \frac{\sqrt{3}}{2} \end{pmatrix}.
\end{equation*}

The interband Berry connection between the lower and upper bands can be expressed as
\begin{equation}
    \vec{A}_{12}(\vec{q}) = \frac{1}{2 |g(\vec{q})|^2} \sum_j{C_j(\vec{q}) \vec{a}_j},
\end{equation}
where $C_j(\vec{q}) = \sum_i \cos \left( \vec{q} \cdot (\vec{a}_i - \vec{a}_j) \right)$.

The condition for the excitation rate to vanish is found by setting the matrix element $|\vec{\epsilon}(\theta_\mathrm{PM}) \cdot \vec{A}_{12}(\vec{q})| = 0$, where
$\vec{q} = q (\cos(\phi) \vec{e}_x + \sin(\phi) \vec{e}_y)$. To lowest order in $q$, we find $\cos(\theta_\mathrm{PM} + 2 \phi) = 0$, giving four distinct solutions
\begin{equation}
    \phi = \left( \frac{m}{2} + \frac{1}{4} \right) \pi - \frac{\theta_\mathrm{PM}}{2} \quad
    (m \in \mathds{Z}).
\end{equation}
For large $q$, higher order terms modify the position of zeroes, but a numerical calculation based on plane wave expansion shows that the deviation is only around $\pi / 30$ at the value of $q$ we analyze, see Fig.~\ref{fig:2:angle_scan}e in the main text.

\subsection{Plane wave expansion}\label{sec:plane_wave}
The Hamiltonian for the stationary system $\hat{H}_0 = \hat{H}_{\text{kin}} + V(\hat{\vec{r}})$ can be expressed in a plane wave basis. We keep plane waves with wave vectors up to $G_\text{max}=5 \vec{G}$, where $\vec{G}$ is a primitive lattice vector. The Hamiltonian is diagonalized numerically and the eigenstates and eigenenergies at all quasimomenta are extracted. 
For Fig.~\ref{fig:3:Full_14_interband_Berry_connection} we write the interband Berry connection in terms of the velocity operator \cite{esteve-paredes_comprehensive_2023} and use
\begin{align}
    \bra{n \vec{q}\,} \vec{v} \ket{n' \vec{q}\,'} &= \frac{1}{m} \sum_{\vec{r}, \vec{r}'} \braket{n \vec{q}\, | \vec{r}} \bra{\vec{r}} \hat{\vec{p}} \ket{\vec{r}\,'} \braket{\vec{r}\,'|n' \vec{q}\,'} \\
            &= \frac{\hbar}{m} \sum_{\vec{G}} (c^{(n, \vec{q})}_{\vec{G}})^{*} \, (\vec{q} + \vec{G}) c^{(n', \vec{q})}_{\vec{G}},\\
            &= \frac{\hbar}{m} \sum_{\vec{G}} (c^{(n, \vec{q})}_{\vec{G}})^{*} \, \vec{G} \, c^{(n', \vec{q})}_{\vec{G}},
\end{align}
where $\braket{\vec{r}\, | n \vec{q}\,} = e^{i\vec{q}\vec{r}} \braket{\vec{r}\, | u^{(n,\vec{q})}} = e^{i \vec{q} \vec{r}} \sum_{\vec{G}} c^{(n, \vec{q})}_{\vec{G}} \, e^{i \vec{G} \vec{r}}$ and we have used Bloch state orthogonality in the last step. 

\subsection{Interpretation of diffraction imaging data}\label{sec:diff_and_BM}

By abruptly turning off the lattice before imaging, we perform a projective measurement in the momentum state basis. As mentioned in the main text, the measured values of $D_{n n'}(\vec{q})$ are smaller than the excitation probabilities $P_{n n'}(\vec{q})$, because the momentum distribution of the $n = 1$ state is not entirely within the first Brillouin zone, and similarly the excited state in band $n'$ may have nonzero amplitude in the first Brillouin zone. Quantitatively, let us consider two Bloch states (dropping $\vec{q}$ dependence for brevity)
\begin{equation*}
    \ket{u_n} = \sum_{\vec{G}} a_{\vec{G}} \ket{\vec{G}},\ \ket{u_{n'}} = \sum_{\vec{G}} b_{\vec{G}} \ket{\vec{G}},
\end{equation*}
where $\ket{\vec{G}}$ is the plane wave state with wave vector $\vec{G}$, and we make the phase choice so that $a_{\vec{0}}$ and $b_{\vec{0}}$ are real positive numbers for this calculation. Then the measured fractional change in column densities $(\rho_{\mathrm{ref}} - \rho_{\mathrm{shake}})/\rho_\mathrm{ref}$ is systematically smaller than the actual fractional decrease in the population in band $n$ by a factor of $a_{\vec{0}}^2 - b_{\vec{0}}^2$.

In principle, since the excitation between $\ket{u_n}$ and $\ket{u_{n'}}$ is driven coherently, the relative phase between the two states should also affect our measurement results. We show that this coherent contribution is canceled out by our measurement protocol. More explicitly, we are interested in measuring the $\ket{\vec{0}}$ population of the state
\begin{equation*}
    \ket{\Psi} = A \ket{u_n} \pm B e^{i \varphi} \ket{u_{n'}},
\end{equation*}
where $A$ and $B$ are positive real numbers, and $B = \sqrt{1 - A^2}$. $+$ and $-$ signs correspond to shaking along $\theta_{\mathrm{PM}}$ and $\theta_{\mathrm{PM}} + \pi$, respectively. In our sequence they correspond to the same polarization, but the modulations start at opposite phases. We have 
\begin{equation}
    \left| \braket{\vec{0} | \Psi} \right|^2 = a_{\vec{0}}^2 \left( A^2 + r^2 B^2 \pm 2rAB \cos{\varphi} \right),
\end{equation}
where $r = b_{\vec{0}} / a_{\vec{0}}$. The last term results from the coherence between $\ket{u_n}$ and $\ket{u_{n'}}$. Averaging the $\theta_{\text{PM}}$ and $\theta_{\text{PM}} + \pi$ image then cancels the coherence term.

\subsection{Details on Data Analysis for Fig.~\ref{fig:2:angle_scan}}\label{sec:angle_scan_analysis}
The trial function used for fitting the angular integral of the annular region shown in Fig.~\ref{fig:2:angle_scan}d-f is given by
\begin{equation*}
    w(\phi) = A + \sum_{i = 1}^4 B_i \exp \left[ \frac{\operatorname{mod}_{(-\pi, \pi]}[\phi - \phi_i]^2}{2 \sigma_i^2} \right]
\end{equation*}
where $\phi$ is the polar angle shown in Fig.~\ref{fig:2:angle_scan}d. $\operatorname{mod}_{(-\pi, \pi]}[\phi - \phi_i]$ means that we take the modulo of $\phi - \phi_i$ such that $-\pi < \phi - \phi_i \leq \pi$ for each $\phi_i$.
The angular positions of the transparency lines are then taken to be $\{ \phi_1, \phi_2, \phi_3, \phi_4 \}$.

\subsection{Extraction of the interband Berry connection}
After collecting data and reference images, we extract the center of mass of the unshaken reference images from atoms in a static lattice and we use it as the $\vec{q} = \Gamma$ position. The length of reciprocal lattice vectors are calibrated using diffraction images of $^{87}\text{Rb}$ Bose-Einstein condensates. Data of the same shaking direction and frequency are averaged over 2 - 6 runs.
\begin{figure}
    \centering
    \includegraphics[width=90mm]{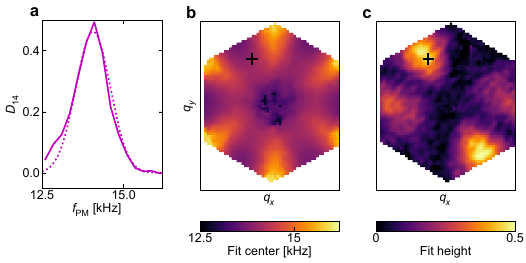}
    \caption{\textbf{Fitting the excitations from diagonal ($\theta_\mathrm{PM} = 45\degree$) linear shaking data.}
    \textbf{a} Excitation spectrum at a single pixel marked in \textbf{b} and \textbf{c}, and in Extended Data Fig.~\ref{fig:FigA_diffraction_panels}. A Gaussian fit (dashed line) to the excitation spectrum (solid line) extracted from experiment reveals the resonant excitation frequency at one quasimomentum. 
    \textbf{b} Performing such fits for each pixel in our images determines the resonant excitation frequency for transitions between bands across the entire Brillouin zone.
    \textbf{c} The extracted Gaussian fit height over the entire first Brillouin zone. The general structure is governed by the orientation and the strength of the interband Berry connection.}
    \label{fig:FgA_maps_plus_spectrum}
\end{figure}

\begin{figure}
    \centering
    \includegraphics[width=90mm]{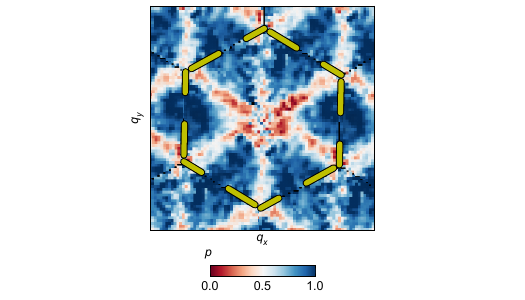}
    \caption{\textbf{$p$ for $\vec{A}_{14}$ extracted from a four parameter fit.}
    Yellow dashed lines mark the edge of the first Brillouin zone.}
    \label{fig:FigA_p_A14}
\end{figure}
We choose a fit function that incorporates all six shaking directions, and we fit the $D_{n n'}$ response vs.\ frequency to six Gaussians that share the same resonance frequency, but have different heights and widths, using a least squares fit. As an example, we show the averaged data in Extended Data Fig.~\ref{fig:FigA_diffraction_panels}, which are fitted to get the results shown in Extended Data Fig.~\ref{fig:FgA_maps_plus_spectrum}. Extended Data Fig.~\ref{fig:FgA_maps_plus_spectrum}a shows an example pixel fit for a diagonal shaking case with an excitation from band 1 to 4. Extended Data Fig.~\ref{fig:FgA_maps_plus_spectrum}b and c show the fit center frequency, which can be used to extract energy band differences, and the fit heights, respectively, for the diagonal linear shaking over the entire Brillouin zone. The fitted peak heights for different shaking directions are used in another least squares fit that fits the data to the expected component of $\vec{A}$ according to Eq.~\ref{eq:rate}. 

We examine whether the experimentally measured $D_{n n'}$ is explained by its relation with the interband Berry connection, as given in Eq.~\ref{eq:rate}, using a generalized set of fit parameters. Under the requirement that the probability of excitation must be real, the most general form of $P_{n n'}$ can be written as $P_{n n'}(\vec{q}) \propto \vec{\epsilon}^* \mathcal{A} \vec{\epsilon}$, where
\begin{align}\label{eqn:general_A}
    \mathcal{A} = \begin{pmatrix}
        \mathcal{A}_{00} & \text{Re}(\mathcal{A}_{01}) + i \, \text{Im}(\mathcal{A}_{01}) \\
        \text{Re}(\mathcal{A}_{01}) - i \, \text{Im}(\mathcal{A}_{01}) & \mathcal{A}_{11} \\
    \end{pmatrix}
\end{align}
In general, $\Re(\mathcal{A}_{01})$ and $\Im(\mathcal{A}_{01})$ are independent of $\mathcal{A}_{00}$ and $\mathcal{A}_{11}$. Using Eq.~\ref{eq:rate} for analysis, as followed in the main text, is equivalent to taking $\mathcal{A} = \vec{A} \vec{A}^{\dagger}$, which contains only three independent parameters, since $\det(\mathcal{A}) = 0$. As explained in the main text, due to hole diffusion and finite momentum resolution, the measured $D_{n n'}$ at each pixel is slightly affected by the population depletion at neighboring pixels. As a result, we may find $\det(\mathcal{A}) > 0$ if the phase of the cross term is not constant in the vicinity of a pixel.

We analyze our data using a four parameter fit defined by Eq.~\ref{eqn:general_A}, and define $p = \sqrt{1 - 4(\det(\mathcal{A})/(\textrm{Tr}(\mathcal{A}))^2} = 1$ as a measure of the validity of Eq.~\ref{eq:rate}. $p = 1$ indicates that $\mathcal{A} = \vec{A} \vec{A}^\dagger$. The result is shown in Extended Data Fig.~\ref{fig:FigA_p_A14}. A three parameter fit well describes the system in regions where $p \approx 1$. We find that $p \not \approx 1$ at quasimomenta where the amplitudes of the cross term are small.

\begin{figure}
    \centering
    \includegraphics[width=90mm]{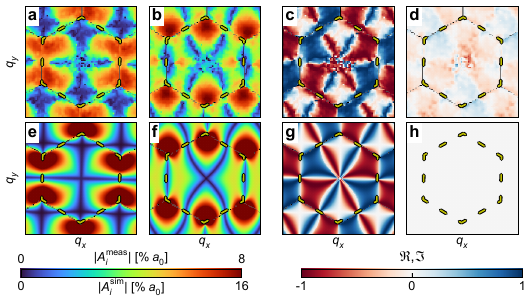}
    \caption{\textbf{Interband Berry connection $\vec{A}_{13}$ shown in a similar fashion as Fig.~\ref{fig:3:Full_14_interband_Berry_connection}.}
    }
    \label{fig:FigA_A13}
\end{figure}

\bibliography{references}

\end{document}